\documentclass[]{aastex631}
\usepackage{aas_macros}
\usepackage{amsmath}
\DeclareMathOperator{\sech}{sech}

\begin{document}

\title{Plasma Dynamics and Nonthermal Particle Acceleration in 3D Nonrelativistic Magnetic Reconnection}

\author{Qile Zhang}
\affiliation{Los Alamos National Laboratory, Los Alamos, NM 87545, USA}
\email{qlzhanggo@gmail.com}

\author{Fan Guo}
\affiliation{Los Alamos National Laboratory, Los Alamos, NM 87545, USA}
\email{guofan@lanl.gov}

\author{William Daughton}
\affiliation{Los Alamos National Laboratory, Los Alamos, NM 87545, USA}
\email{daughton@lanl.gov}

\author{Xiaocan Li}
\affiliation{Dartmouth College, Hanover, NH 03755, USA}
\email{phyxiaolee@gmail.com}

\author{Hui Li}
\affiliation{Los Alamos National Laboratory, Los Alamos, NM 87545, USA}
\email{hli@lanl.gov}



\begin{abstract}
Understanding plasma dynamics and nonthermal particle acceleration in 3D magnetic reconnection has been a long-standing challenge. In this paper, we explore these problems by performing large-scale fully kinetic simulations of multi-xline plasmoid reconnection with various parameters in both the weak and strong guide field regimes. In each regime, we have identified its unique 3D dynamics that leads to field-line chaos and efficient acceleration, and we have achieved nonthermal acceleration of both electrons and protons into power-law spectra. The spectral indices agree well with a simple Fermi acceleration theory that includes guide field dependence. In the low-guide-field regime, the flux-rope kink instability governs the 3D dynamics for efficient acceleration. The weak dependence of the spectra on the ion-to-electron mass ratio and $\beta$ ($\ll1$) implies that the particles are sufficiently magnetized for Fermi acceleration in our simulations. While both electrons and protons are injected at reconnection exhausts, protons are primarily injected by perpendicular electric fields through Fermi reflections and electrons are injected by a combination of perpendicular and parallel electric fields. The magnetic power spectra agree with in-situ magnetotail observations, and the spectral index may reflect a reconnection-driven size distribution of plasmoids instead of Goldreich–Sridhar vortex cascade. As the guide field becomes stronger, the oblique flux ropes of large sizes capture the main 3D dynamics for efficient acceleration. Intriguingly, the oblique flux ropes can also run into flux-rope kink instability to drive extra 3D dynamics. This work has broad implications for 3D reconnection dynamics and particle acceleration in heliophysics and astrophysics.

\end{abstract}



\section{Introduction} \label{sec:intro}
Magnetic reconnection is a fundamental physics process in magnetized plasmas that releases magnetic energy and drives particle acceleration in various energetic phenomena in space and astrophysics \citep{Yamada2010}. Energetic particles are often observed during magnetic reconnection in space and solar plasmas -- for example, in Earth's magnetotail \citep{Ergun2020,Oka2023SSRv}, the heliospheric current sheets \citep{Desai2022,Zhang2024prl}, coronal interchange reconnection \citep{Bale2023nature} and solar flares \citep{Gary2018,Chen2020} with low plasma $\beta$. Magnetic reconnection is observed to accelerate electrons \citep{Krucker2010,Oka2015,Oka2018,Gary2018,Lin2011}, protons \citep{Omodei2018,Cohen2020,Bale2023nature} and heavier ions \citep{Desai2022,Cohen2020,Bale2023nature} into nonthermal power-law energy distributions $f \propto E^{-p}$ (with a wide range of spectral indices $p$ ranging from 3 to 9). The nonthermal acceleration often occurs simultaneously for ions and electrons \citep{Shih2009,Ergun2018,Ergun2020}.  This indicates a common acceleration process in low-$\beta$ reconnection for both ions and electrons. However, the underlying mechanisms have been challenging to understand since most previous studies failed to produce the simultaneous ion and electron power laws in self-consistent kinetic reconnection simulations \citep{Dahlin2014,Dahlin2017,Li2017,Li2018,Li2019b,Li2021review,Guo2020review}.  Meanwhile, during acceleration reconnection also drives a turbulent state as seen in the magnetotail \citep{Ergun2018,Ergun2020} and solar flares \citep{Cheng2018,French2019}. Unfortunately, the relation between reconnection and the turbulence is also unclear.
% \textbf{Review some hard X-ray papers showing electron power-law has different spectral index, also perhaps some observation evidence for protons and heavy ions. Also perhaps you can also discuss the Ergun paper which suggests the roles of turbulence.}

%\textbf{add a paragraph introducing the basic acceleration mechanisms in magnetic reconnection. What's the main unknowns (power-law formation, 3D turbulence, etc.)}
%Fermi acc is found to be the major acc mechanism but how rec form the power laws and 3D turbulence is still unknown. 
Recent studies have found the major acceleration mechanism to be the Fermi acceleration mechanism \citep{Drake2006,Guo_2014,Dahlin2014,Dahlin2016pop,Dahlin2017,Li2017,Li2018,Li2019b}, where particles bounce back and forth off contracting field lines to reach high energy. Since Fermi acceleration rate is proportional to the particle energy, at high energy it will over run the acceleration by the parallel electric field  \citep{Dahlin2016pop,Zhang2019,Zhang2019b,Le2009,Haggerty2015}.  Since the Fermi mechanism is driven by field line curvature and the curvature is strongest in the low guide field regime, this mechanism is most efficient with a low guide field \citep{Dahlin2016pop,Li2017,Li2018,Li2019b,Arnold2021}. The energy gain is strongest in the weak guide field regime and weaker for higher guide fields. However, magnetic islands in 2D can trap energetic electrons to prevent further Fermi acceleration \citep{Dahlin2014,Li2017,Johnson2022}. Previous studies show that 3D turbulent dynamics that produces field-line chaos can facilitate electron transport out of magnetic islands towards acceleration regions (reconnection exhausts) for more efficient Fermi acceleration \citep{Dahlin2017,Li2019b}. This stronger acceleration in 3D can help to form sustainable power laws for electrons \citep{Li2019b}. However, it is still not clear how ions and electrons can be both accelerated and develop power-law energy spectra, as observations indicated, and what are the origin and nature of the turbulent state in the reconnection region, as will be further discussed below.

Recently, \cite{Zhang2021} for the first time produce simultaneous ion and electron power laws in fully kinetic 3D simulations in the low-guide-field regime ($b_g<0.5$). 
% In this regime the major acceleration mechanism -- Fermi acceleration (where particles bounce off contracting field lines) \citep{Drake2006,Dahlin2014,Dahlin2016,Dahlin2017,Li2017,Li2018,Li2019b} -- is most efficient due to the strongest magnetic curvature and tension \citep{Dahlin2016,Li2017,Li2018,Li2019b,Arnold2021}.
They achieve these power laws by taking advantage of the domain-size threshold for the flux-rope kink instability (of the tearing-mode generated magnetic flux ropes) \citep{Zhang2021,Dahlburg1992} in the 3D domain design. This instability can disrupt and fragmentize the flux ropes (see also \cite{Zhang2024prl}) to turn the reconnection layer into a turbulent state. This controls the 3D field-line chaos that facilitates efficient Fermi acceleration. Note that this instability is distinct from the more recognized drift-kink instability \citep{Daughton1998,Zenitani2005,WLiu2011} where the current sheet flaps. This study \citep{Zhang2021} creates new opportunities to study 3D reconnection dynamics, as well as ion and electron acceleration in fully kinetic 3D simulations. In this paper we use these simulations to further explore the important aspects such as 3D dynamics, parameter dependence, injection physics and magnetic power spectra.

While the Fermi acceleration is most efficient in the low-guide-field regime, it is still important in the regime with a somewhat higher guide field ($0.5<b_g\lesssim1$). The ion and electron acceleration with such guide fields also commonly occurs in the solar corona, solar flares, solar wind and the magnetosphere. Thus, it is important to achieve and understand the nonthermal ion and electron acceleration in this regime. Previous studies suggested that the 3D turbulence and field-line chaos that facilitate efficient acceleration is driven by the overlapping oblique tearing-mode flux ropes \citep{Bowers2007,Daughton2011,Liu2013prl,Onofri2006}. However, the tearing modes from kinetic reconnection current sheets are at kinetic scales, which are usually much smaller than the system size. While earlier studies have recognized their chaotic and turbulent nature \citep{Daughton2011,Liu2013prl,Dahlin2017}, it is not completely clear what happens after these kinetic-size flux ropes continue to grow as reconnection proceeds.
%But to fully capture the 3D effect, it is unclear flux ropes of what sizes matter: whether it is the oblique flux ropes at the tearing mode instability, with sizes of current sheet thickness, or it is the oblique flux ropes that keep growing when advecting with reconnection outflows, with large sizes eventually proportional to the system size. These two types of flux ropes require drastically different domain sizes in the guide field direction to be contained in the periodic simulation domain. 
Moreover, in the light of the flux-rope kink instability for low guide fields, the oblique flux ropes could also be subject to the kink instability. It is thus unclear whether the overlapping oblique flux ropes are the only important process in the 3D turbulent dynamics with a guide field. Therefore, in this paper we will explore the higher-guide-field regime  regarding the 3D dynamics and nonthermal acceleration for ions and electrons.

This paper is arranged as follows. In Section \ref{sec:lowbg}, after we demonstrate nonthermal ion and electron acceleration in the low-guide-field regime using 3D fully kinetic particle-in-cell (PIC) simulations, we explore additional important aspects such as plasma $\beta$ and ion-to-electron mass ratio dependence, injection process and magnetic power spectra, to gain further insight into the acceleration process. We find that the magnetic power spectra agree well with in-situ magnetotail observations, and the spectral indices may reflect a reconnection-driven size distribution of magnetic flux ropes (or islands) instead of Goldreich–Sridhar vortex cascade. Then in Section \ref{sec:higherBg} we switch to the higher-guide-field regime. We find in our simulations that the initially small flux ropes from oblique tearing modes can keep on growing while maintaining their oblique angles and advecting with the large bidirectional reconnection outflows -- eventually becoming large and proportional to the system size. It is the flux ropes of large sizes (proportional to the system size) that control the domain-size threshold to capture the 3D field-line chaos for efficient acceleration. By taking advantage of this threshold, for the first time our 3D PIC simulations with higher guide fields accelerate both ions and electrons into power-law energy spectra. The power-law indices are consistent with the Fermi-acceleration predictions, with steeper spectral slopes than the low-guide-field regime due to the weaker acceleration. We discover that the oblique flux ropes can also be kink unstable -- which gives rise to another new domain-size threshold -- driving extra 3D dynamics to the reconnection layer. However, this oblique flux-rope kink instability (and its driven 3D dynamics) does not appear to significantly further enhance the acceleration for ions and electrons in our simulations. These results have broad applications for particle acceleration by reconnection in the magnetosphere, solar wind and solar corona.

\section{simulation setup} \label{sec:style}
We use the VPIC code that solves the Vlasov-Maxwell equations \citep{Bowers2008}. The 3D simulations start from a force-free layer $\mathbf{B}=B_0 \tanh(z/\lambda)\mathbf{e_x}+\sqrt{B_0^2 \sech^2(z/\lambda)+B_g^2}\mathbf{e_y}$ with a uniform plasma density $n_i = n_e = n_0$. $B_0$ is the reconnecting field, $B_g$ is the guide field and $\lambda$ is the half-thickness of the layer, which is set to be one ion inertial length $d_i$. Electrons carry the initial current that satisfies the Ampere's law. The ratio of plasma frequency to electron cyclotron frequency $\omega_{pe}/\Omega_{ce}$ is set to be 1. The default grid size $\Delta x=\Delta y=\Delta z=0.0488d_i$ (which changes in some simulations, as shown in Table \ref{table1}), with 150 particles per cell for each species. 
%{Most simulations have proton-to-electron mass ratio $m_i/m_e=25$, $b_g=B_g/B_0=0.2$ and $V_A=B_0/\sqrt{4\pi n_0m_i}=0.2c$, where $c$ is the speed of light. The initial temperature $T_i=T_e=0.01m_iV_A^2$ so the plasma $\beta$ based on the reconnecting field $\beta=0.02$. The grid size is $\Delta x=\Delta y=\Delta z=0.0488d_i$, with 150 particles per cell per species.}
Boundary conditions are periodic in $x$ and $y$, and conducting for fields  and reflecting for particles in $z$. A small long-wavelength perturbation with $B_z=0.02B_0$ is included to initiate reconnection. To limit the influence of periodic boundaries, unless specified, all simulations terminate at about $1.3$ Alfv\'en crossing time $L_x/V_A$ before the acceleration stagnates. A set of simulations have been conducted to study the underlying processes for different guide fields (from 0.2 to 1), electron/proton $\beta$ based on reconnecting fields (from 0.02 to 0.08), domain sizes ($L_x = 75 - 300 d_i$), and the mass ratio ($m_i/m_e = $25 or 100 with corresponding $c/V_A=5,10$). These simulations are summarized in Table \ref{table1}. The ``field-line chaos'' in Table \ref{table1} indicates whether the simulations show 3D effects with chaotic field lines. 

\begin{widetext}
\begin{center}
\begin{table}
\begin{tabular}{  c c c c c c c c c c  } 
 \hline
Run	& $L_x/d_i$	& $L_y/d_i$	& $L_z/d_i$ &  $\Delta x/d_i$ & $\beta_{xe}$ &$B_g/B_0$&$m_i/m_e$&3D field-line chaos\\
 \hline
1	& 150	& 6.25 & 	62.5 & 0.0488&	0.02 &0.2&25 &	No \\
2	& 150	& 12.5 & 62.5 & 0.0488&	0.02 &0.2&25&  	 Yes\\
3	& 300	& 25.0 & 125.0 &0.0488&0.02 &0.2&25 & 	Yes \\
4	& 150	& 12.5 & 62.5 &0.0488& 	0.08 &0.2&25&  	Yes \\
5	& 75	& 6.25 & 31.25 &0.0488& 	0.02 &0.2&25& 	Yes\\
6	& 75	& 6.25 & 31.25 &0.0244& 	0.02 &0.2&100& 	Yes\\
7	& 150	& 15.625 & 	62.5 &0.0488& 	0.02 &0.6&25  & 	Yes \\
8	& 150	& 6.25 & 62.5 &0.0488& 	0.02 &0.6&25&  	 No\\
9	& 150	& 25 & 	62.5 &0.0488& 	0.02 &0.6&25  & 	Yes \\
10	& 150	& 75 & 62.5 &0.0488& 	0.02 &0.6&25&  	 Yes\\
11	& 300	& 50.0 & 125.0 &0.061& 0.03125 &0.6&25 &  	Yes \\
12	& 300	& 50.0 & 125.0 &0.061& 0.03125 &1.0&25 &  	Yes \\
\hline
\end{tabular}
\caption{Simulations discussed in this paper. $\beta_{xe}$ means electron $\beta$ based on reconnecting fields. The the last column ``3D field-line chaos'' indicates whether the simulation shows 3D dynamics with chaotic field lines.\label{table1}}
\end{table}
\end{center}
\end{widetext}

\section{reconnection with a low guide field} \label{sec:lowbg}
\subsection{3D Dynamics}
\begin{figure*}
\begin{center}

\includegraphics[width=0.6\textwidth]{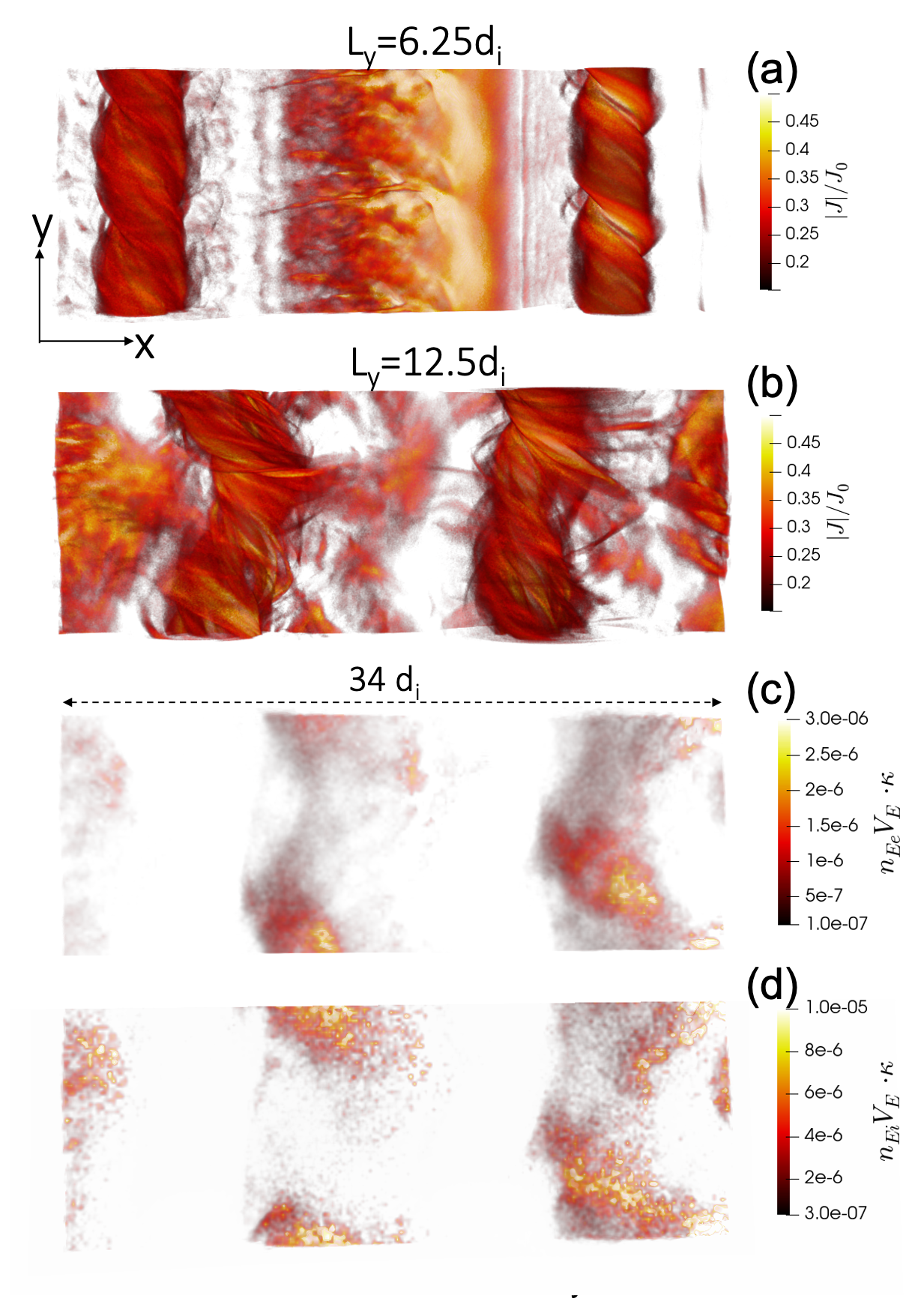} %figE.2.png}
\caption{Current density $|J|/en_0c$ for simulations Run 1 and 2 with different $y$ dimensions (a) $L_y=6.25d_i$ and (b) $L_y=12.5d_i$ respectively, at $t\Omega_{ci}=200$. (c) and (d) show the energetic electron and proton density ($n_{Ee}$ and $n_{Ei}$ with energy $1.2 < \varepsilon/m_iV_A^2 < 2.4$) multiplied by $V_E \cdot \kappa$, where $V_E$ is the $E\times B$ flow and $\kappa$ is the magnetic curvature vector.\label{fig1}}
\end{center}
\end{figure*}

%m=1 flux-rope kink tears up the flux surface and make the layer turbulent.
In the low-guide-field regime, we find that the $m=1$ flux-rope kink instability drives the turbulent 3D dynamics by disrupting the flux ropes \citep{Zhang2021}. This belongs to the ``external kink instability'' in plasma physics. The flux ropes only become $m=1$ kink unstable when its length is long enough to make the safety factor at the edge of flux ropes
\begin{equation}
    q_{edge}\sim\pi b_gD/L_y<1
\end{equation}
\citep{Oz2011} (the``Kruskal–Shafranov limit"), where $D$ is the typical diameter of the flux ropes. This means that the instability of flux ropes takes place when $L_y$ exceeds a threshold 
\begin{equation}
    L_{th}\sim \pi b_g D\sim 0.1\pi b_g L_x,
\end{equation}
given that approximately $D\sim0.1L_x$.
We will demonstrate this below with simulations Run 1 and 2  in Table \ref{table1} with $b_g=0.2, L_x\sim150d_i$, so $L_{th} \sim 9.5d_i$. They have $L_y$ below and above $L_{th}$ to be stable and unstable. %, which divides the $L_y$ of these two stable and unstable simulations.% agreeing well with the simulation results. 
As shown in Figure \ref{fig1}(a) with current density,  the flux ropes in Run 1 with $L_y=6.25d_i<L_{th}$  are stable to $m=1$ kink, which are nearly 2D like and laminar, although higher-harmonic ($m>1$) kink modes may develop. In contrast, in Run 2 with two times larger $L_y=12.5d_i>L_{th}$ and otherwise the same parameters (Figure \ref{fig1}(b)), the flux ropes present $m=1$ flux-rope kink instability, which tear up the otherwise closed flux surfaces and make the reconnection layer turbulent with 3D field-line chaos.

\subsection{Nonthermal Particle Acceleration}
%The field-line chaos and transport out of flux ropes facilitates efficient Fermi acceleration.
The particle transport due to this kink-driven field-line chaos enables the energetic particles to easily access the major Fermi acceleration regions at the reconnection exhausts for more efficient acceleration. Figure \ref{fig1}(c-d) shows the energetic electron and proton densities (with energy $1.2 < \varepsilon/m_iV_A^2 < 2.4$) in the kink unstable simulation Run 2, multiplied by $V_E \cdot \kappa$ ($V_E$ is the $E \times B$ drift velocity and $\kappa$ is the magnetic curvature) that quantifies the field-line contraction and Fermi acceleration rate \citep{Dahlin2017,Li2019b}. The values maximizing at the exhausts adjacent to the flux ropes suggest that the energetic particles overlap with the major Fermi acceleration regions at the exhausts (with strongest magnetic curvature) for efficient acceleration. 

\begin{figure*}
    \centering
\includegraphics[width=0.9\textwidth]{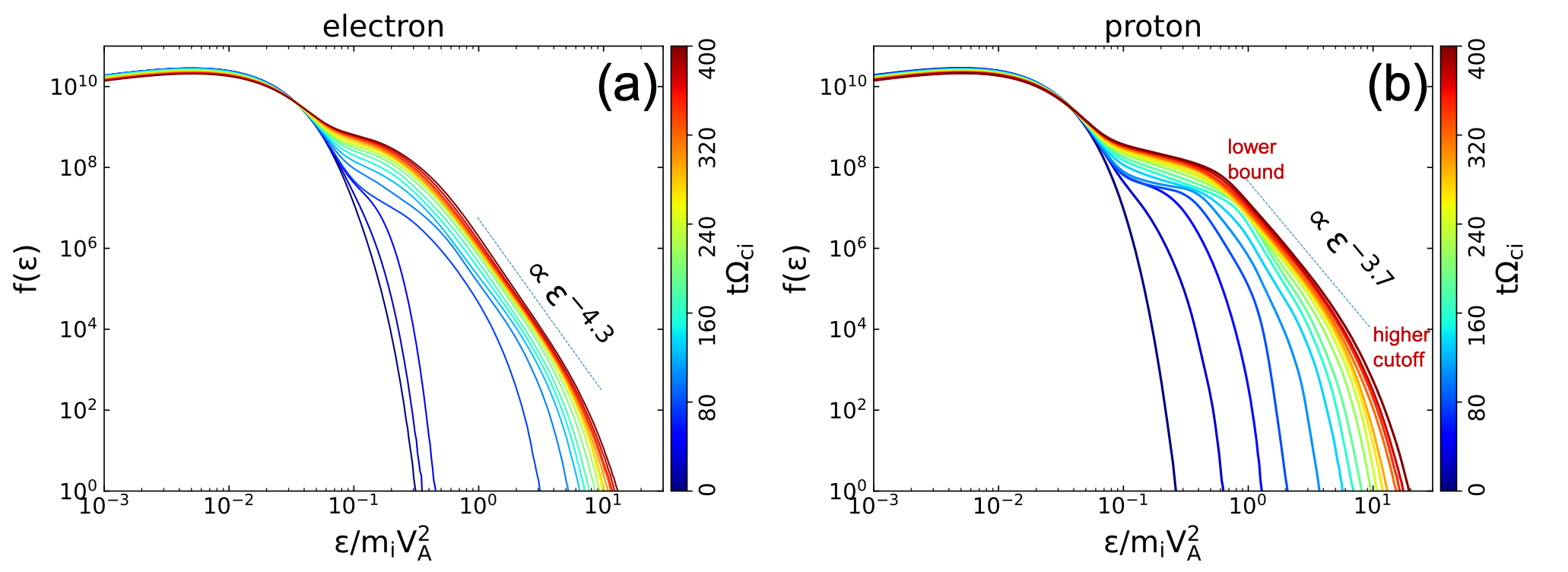} %figE.2.png}
\caption{Evolution of energy spectra for electrons and protons in the large simulation Run 3.\label{fig2}}
\end{figure*}

Since the $m=1$ kink instability controls the efficient acceleration, we take advantage of its $L_y$ threshold  $L_{th}$ and perform a 3D simulation of unprecedented size in $x$ with $L_x=300d_i$ (Run 3 in Table \ref{table1}). As a result, in this 3D simulation with $m=1$ flux-rope kink instability, both electrons and protons are accelerated into clear nonthermal power-law spectra (Figure \ref{fig2}), with indices around 4. The spectra have several distinct features: the low energy bound of the power law (the shoulders, $E_{l,e}\sim0.2m_iV_A^2$, $E_{l,i}\sim0.5m_iV_A^2$)) indicating the injection energy for particles, the power laws formed and extended by the Fermi acceleration process after injection, and the power-law high energy cutoff ($E_{h,e}\sim E_{h,i}\sim7m_iV_A^2$) indicating the maximum energy particles are accelerated to.

The Fermi acceleration process can be described by the a scaling analysis in \cite{Zhang2021}. 
Using particle acceleration theory and considering Fermi acceleration at reconnection exhausts,
%the power law index $p=1+(\alpha \tau_{esc})^{-1}$, where $\alpha$ is the Fermi acceleration rate and $\tau_{esc}^{-1}$ is the escape rate from the reconnection layer. At typical acceleration regions (exhausts) $\alpha\equiv{\dot{\varepsilon}}/\varepsilon\sim \mathbf{U_E}\cdot\mathbf{\kappa}\sim V_{Ax}\kappa_x$ and  $\tau_{esc}^{-1}\sim V_{Ax}/L$, where $\mathbf{U_E}$ is the E$\times$B drift speed, $\mathbf{\kappa}$ is the magnetic field curvature vector and $L$ is the half length of the reconnecting current sheet. 
% we get
% \begin{equation}
% p \sim 1+\frac{V_{Ax}/L}{V_{Ax}\kappa_x}
%  =1+\frac{1}{L \kappa_x},
% \end{equation}
% Since $\kappa_x=(\hat{b}\cdot\nabla\hat{b})_x\sim \hat{b}_z\cdot\partial_z \hat{b}_x
% \sim B_z B_x/(B^2\Delta_z$), 
% %\begin{equation}
% %\kappa_x=(\hat{b}\cdot\nabla\hat{b})_x\sim \hat{b}_z\cdot\partial_z \hat{b}_x
% %\sim B_z B_x/B^2/\Delta_z,
% %\end{equation}
% where $\Delta_z$ is the typical length scale of exhaust field lines in z (related to the scale of flux ropes), 
we obtain

%\begin{equation}
%\kappa_x= \frac{0.05B_x^2}{(B_x^2+B_g^2)\Delta_z}.
%\end{equation}
\begin{equation}
p\sim 1+\frac{B_x}{B_z}\frac{\Delta_z}{L}(1+\frac{B_g^2}{B_x^2}), \label{p_formula}
\end{equation}
where $\Delta_z$ is the typical length scale in $z$ of exhaust field lines (related to the scale of flux ropes), $L$ is the half length of the reconnecting current sheet, and the scales of magnetic fields $B_x$ $B_z$ are evaluated in the acceleration regions (exhausts). 
% \begin{equation}
% \alpha\sim V_{Ax}\kappa_x=\frac{B_z V_{Ax}B_x^2}{B_x(B_x^2+B_g^2)\Delta_z}.
% \end{equation}
Considering $\Delta_z$ and $L$ are both proportional to the domain size, they are roughly proportional to each other and thus the predicted spectral indices remain unchanged for larger domains.
According to the typical values in the exhausts in our low-guide-field simulations, we obtain $p\sim4$, consistent with simulation results as discussed in \citep{Zhang2021}. Equation (\ref{p_formula}) not only applies to the low-guide-field regime but also to the higher-guide-field regime, which will be further discussed below in Section \ref{sec:higherBg}.
\subsection{Parameter Dependence}
\begin{figure*}
\begin{center}
\includegraphics[width=0.75\textwidth]{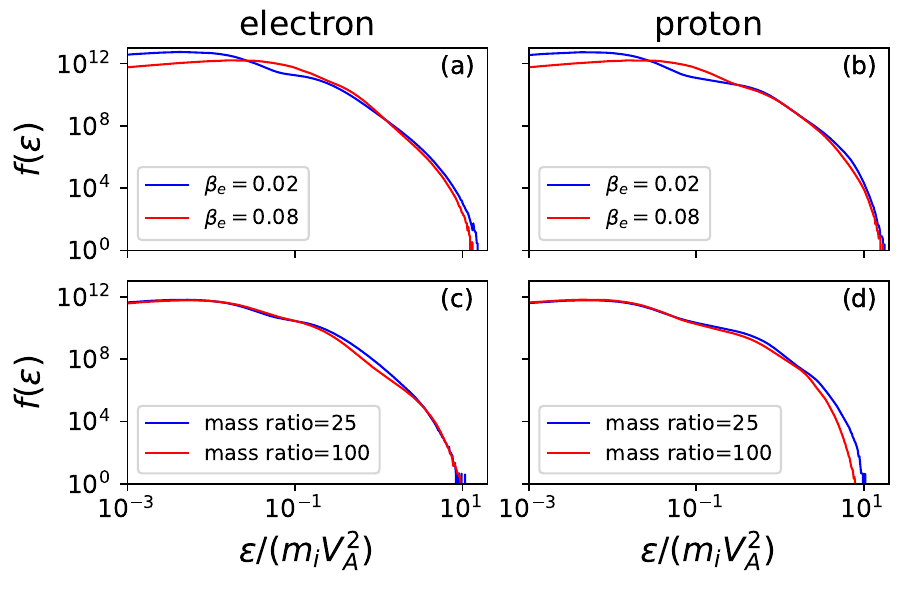} %figE.2.png}
\caption{Dependence of energy spectra for electrons and protons. (a) and (b) show the dependence on $\beta$ in Run 2 and 4 at $t\Omega_{ci}=200$. (c) and (d) show the dependence on mass ratio in run 5 and 6 at $t\Omega_{ci}=125$. \label{fig3}}
\end{center}
\end{figure*}

%The weak dependence of the spectra on mass ratio and beta (¡¡1) implies the plasma are magnetized well enough for nonthermal acceleration in our simulations
Here we examine the dependence of the energy spectra to parameters like plasma $\beta$ (in the low-$\beta$ regime), and ion-to-electron mass ratio. Figure \ref{fig3}(a-b) shows electron and proton spectra with two $\beta$ in the low-$\beta$ regime. Due to the temperature change, the upstream Maxwellian distribution has a significant shift, but the power-law slopes at higher energy remains very similar. Also, the low energy shoulder of the proton spectrum remains essentially unchanged around $0.5m_iV_A^2$ \citep{Zhang2021}. This is because in the low-$\beta$ regime, the magnetic tension and the Fermi reflection process are not sensitive to $\beta$. Different $\beta$ would not significantly affect the power-law slopes from the Fermi acceleration process and the low energy shoulder of protons from the first Fermi reflection process \citep{Zhang2021}, which is consistent with our understanding of Fermi acceleration in reconnection. Figure \ref{fig3}(c-d) show weak dependence of the spectra on the mass ratio, although it is still far away from the realistic mass ratio. A higher mass ratio is essentially reducing the electron mass in the simulation, so the ion acceleration may not substantially change with a higher mass ratio. Interestingly the electron acceleration appears to have almost no change on the maximum energy with the mass ratio (panel (c)). The no extra electron acceleration for higher mass ratio (smaller Larmor radii) suggests that electrons are magnetized enough by magnetic flux ropes and exhausts to continue Fermi acceleration, even with mass ratio 25 -- despite that energetic electrons can be scattered and isotropized by 3D turbulence \citep{Li2019b}. Previous 2D simulations also showed no extra electron acceleration for higher mass ratios \citep{Li2019a}. In fact, the Larmor radii of energetic electrons for mass ratio 25 are still much less than the typical size of flux ropes in simulations. As the particles get accelerated with $\varepsilon\propto t^{0.8}$ \citep{Zhang2021}, their Larmor radii ($\rho\propto\varepsilon^{0.5}$) grow much slower than the growth of flux-rope size ($\propto t$) as reconnection proceeds. Therefore, the energetic particles can actually get more and more magnetized by the flux ropes for further Fermi acceleration. This disagrees with a hypothesis in \cite{Arnold2021} that energetic electrons in PIC simulations have too large Larmor radii and thus get demagnetized to stop Fermi acceleration, suppressing the extension of power laws. The more extended electron power laws in the macroscale model than PIC simulations -- produced by the macroscale reconnection acceleration model kglobal \citep{Arnold2021} -- may result from other important factors such as the much larger effective domain sizes proportional to the acceleration time \citep{Zhang2021,Zhang2024prl}, and the absence of pitch angle scattering that maximizes the parallel energization in favor of Fermi acceleration. This needs to be further explored. 

\subsection{Injection Process}
%While the both electrons and protons can be injected at the reconnection exhaust, protons are majorly injected by perpendicular electric field through Fermi reflection and electrons are partically by perpendicular and parallel electric fields.

\begin{figure*}
    \centering
\includegraphics[width=0.9\textwidth]{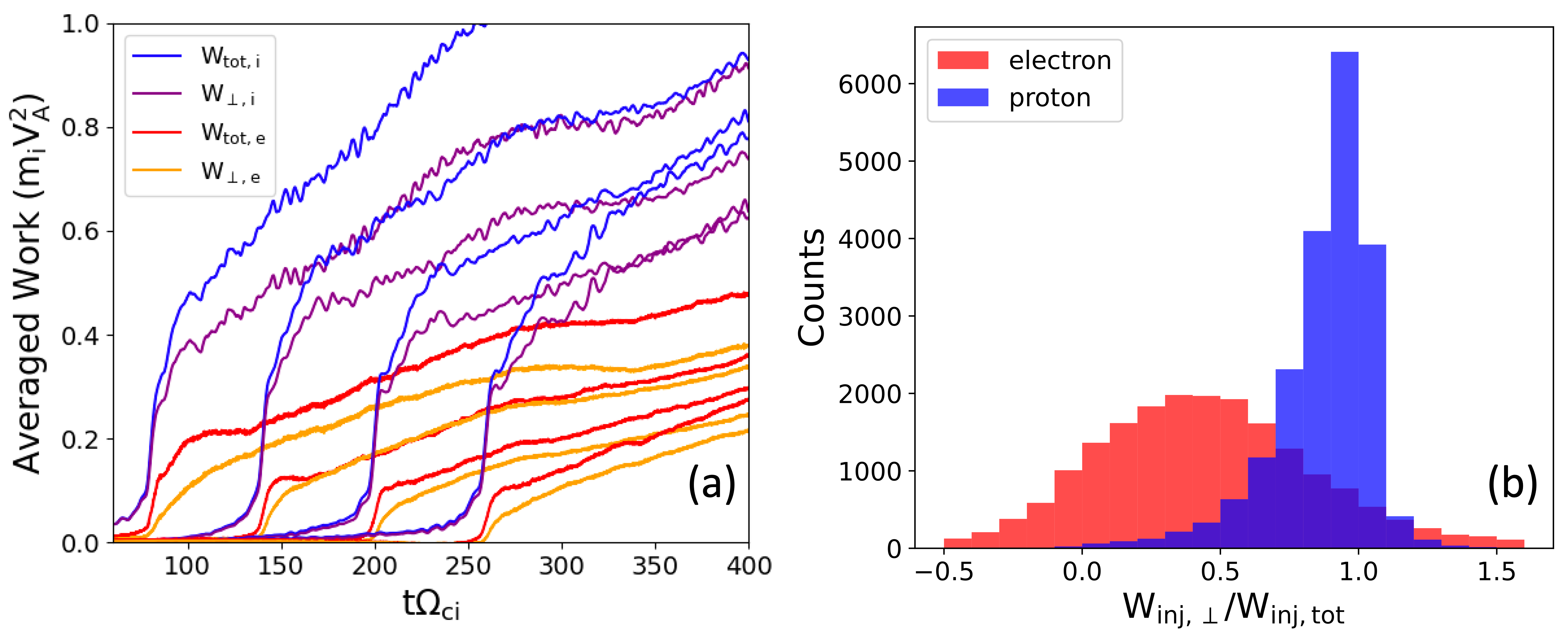} %figE.2.png}
\caption{(a) Evolution of work done by total and the perpendicular component of electric fields for protons and electrons in Run 3, averaged over different generations of injected particles. (b) histogram of the perpendicular-work contribution fraction for injection for both species.  \label{fig4}}
\end{figure*}

Here we explore the relative contribution of perpendicular and parallel electric fields during the particle injection process for both species using Run 3. Figure \ref{fig4}(a) shows the evolution of the total work done and its perpendicular-electric-field component, averaged within each different generation of electrons and protons starting acceleration at different times. Particles are included in a generation if the final energies are
above the spectral low-energy bound and if the starting time of energization
is within an $\Omega_{ci} \Delta t=5$ interval. It shows that the protons are mostly injected by the perpendicular electric field, while the electrons are only halfly injected by it. We also show histogram of injection contribution percentage (Figure \ref{fig4}(b)), suggesting a similar conclusion: the proton perpendicular contribution peaks around 100\% while electrons' peaks around 50\%. 

\begin{figure*}
    \centering
\includegraphics[width=0.9\textwidth]{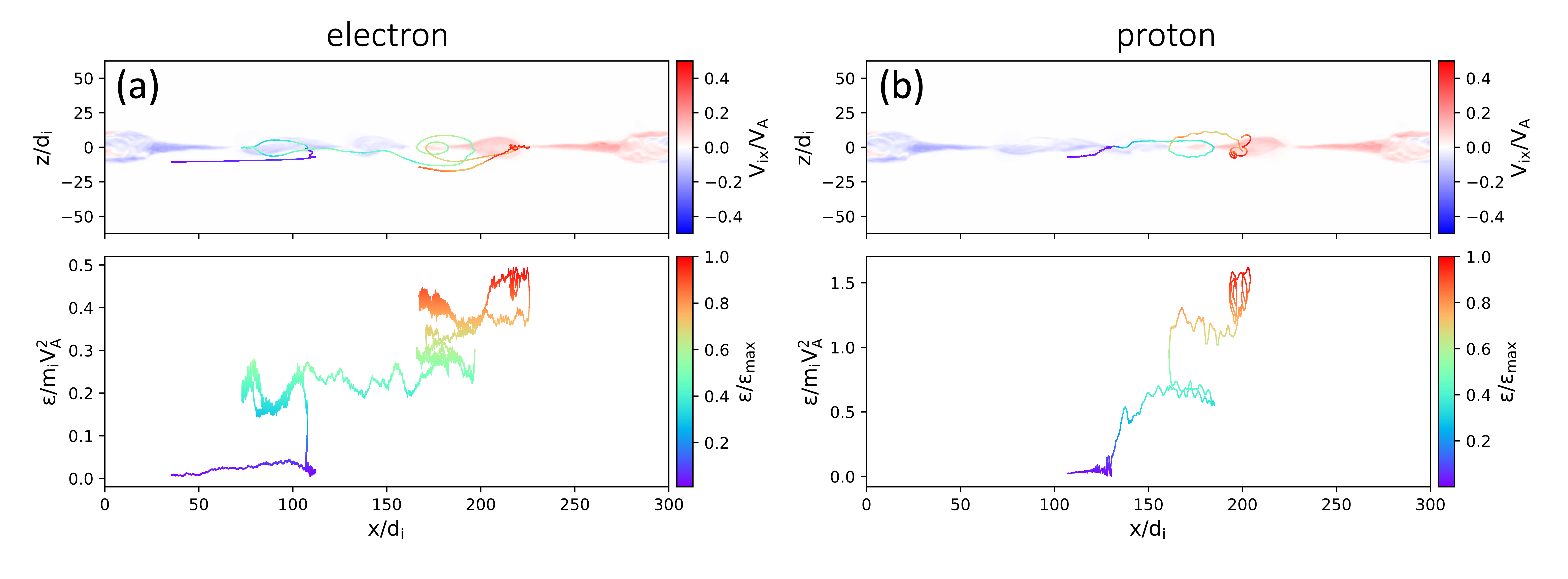} %figE.2.png}
\caption{Representative particle trajectories in Run 3 demonstrating the injection process for both species, colored by the particle energy $\varepsilon$ normalized by the maximum energy $\varepsilon_{max}$. The top panels show proton velocity in $x$ ($V_{ix}$) as backgrounds around the injection time of the particles (a 2D $x-z$ slice at the $y$ location where the particle is injected). \label{fig5}}
\end{figure*}

We also show two representative particles (an electron and a proton) to demonstrate their injection process at reconnection exhausts. Figure \ref{fig5}(a-b) shows the background of $V_{ix}$ around the injection time of the particles, with the particle trajectories (colored by energy) overlaid. When particles for the first time cross an exhaust from upstream, they get boosted to the injection energy (around $0.2m_iV_A^2$ for electrons and $0.5m_iV_A^2$ for protons) as shown Figure \ref{fig5}(c-d). 
%As shown in the electric-field-component contribution above, the protons are injected by a Fermi reflection at the exhaust, and electrons by a combination of Fermi reflection and parallel electric fields. 
After injection, the particles wander elsewhere and get further Fermi acceleration. 

\subsection{Magnetic Power Spectra} \label{sec:powerspectra}
The magnetic power spectra contain important information about the fluctuation energy in the reconnection layer, and they are measured in both spacecraft observations and reconnection simulations. However, the magnetic power spectra do not appear to agree well between observations and simulations.  MMS along the turbulent reconnection layer at the magnetotail \citep{Ergun2018,Ergun2020} observe magnetic power spectra with indices $\alpha\sim5/3$ and 3 below and above $k d_i\sim1$. The observed frequent change of $B_z$ direction indicates that it may be in the plasmoid reconnection regime relevant to our simulations. A recent paper \citep{Richard2024prl} performed a statistical analysis over 24 reconnection jets and found an averaged power spectrum with similar indices as above, but with a shorter inertial range. Each individual event has a variety of power spectrum (see their Fig. 2a) and is harder to compare with. The $\alpha\sim5/3$ index at super-ion scales is often suggested to be a signature of the inertial range of a Goldreich–Sridhar turbulence cascade, in which theory the vortex energy cascades across scales and form spectra over $k_\perp$ with index $\alpha=5/3$. Here the $k_\perp$ describes perturbations perpendicular to a dominant uniform mean magnetic field. 
But reconnection at the magnetotail does not have a dominant uniform mean field, with significant field changes ($\Delta B/B\sim1$) across the reconnection midplane and flux ropes. In fact, the significant field changes within the inertial range scales are frequently observed in \cite{Ergun2018,Ergun2020}. Thus, it does not have uniform parallel and perpendicular directions and it is not straightforward that the Goldreich–Sridhar cascade is applicable here.
If we stick to this framework and average the fields over the whole reconnection region, the overall mean field is the guide field and one can calculate  $k_\perp$ spectra perpendicular to the guide field from simulations. 
Large-scale 3D fully kinetic simulations in the plasmoid reconnection regime have reconnection driven 3D turbulence relevant to the observed turbulent reconnection \citep{Li2019b,Daughton2014pop,Guo2021apj}. 
In these simulations, the $k_\perp$ spectra have index about 2.7, which does not match the 5/3 in observations. 
Therefore, it is not obvious that the 5/3 index in observations corresponds to the Goldreich–Sridhar vortex cascade.  

\begin{figure*}
\centering
\includegraphics[width=0.85\textwidth]{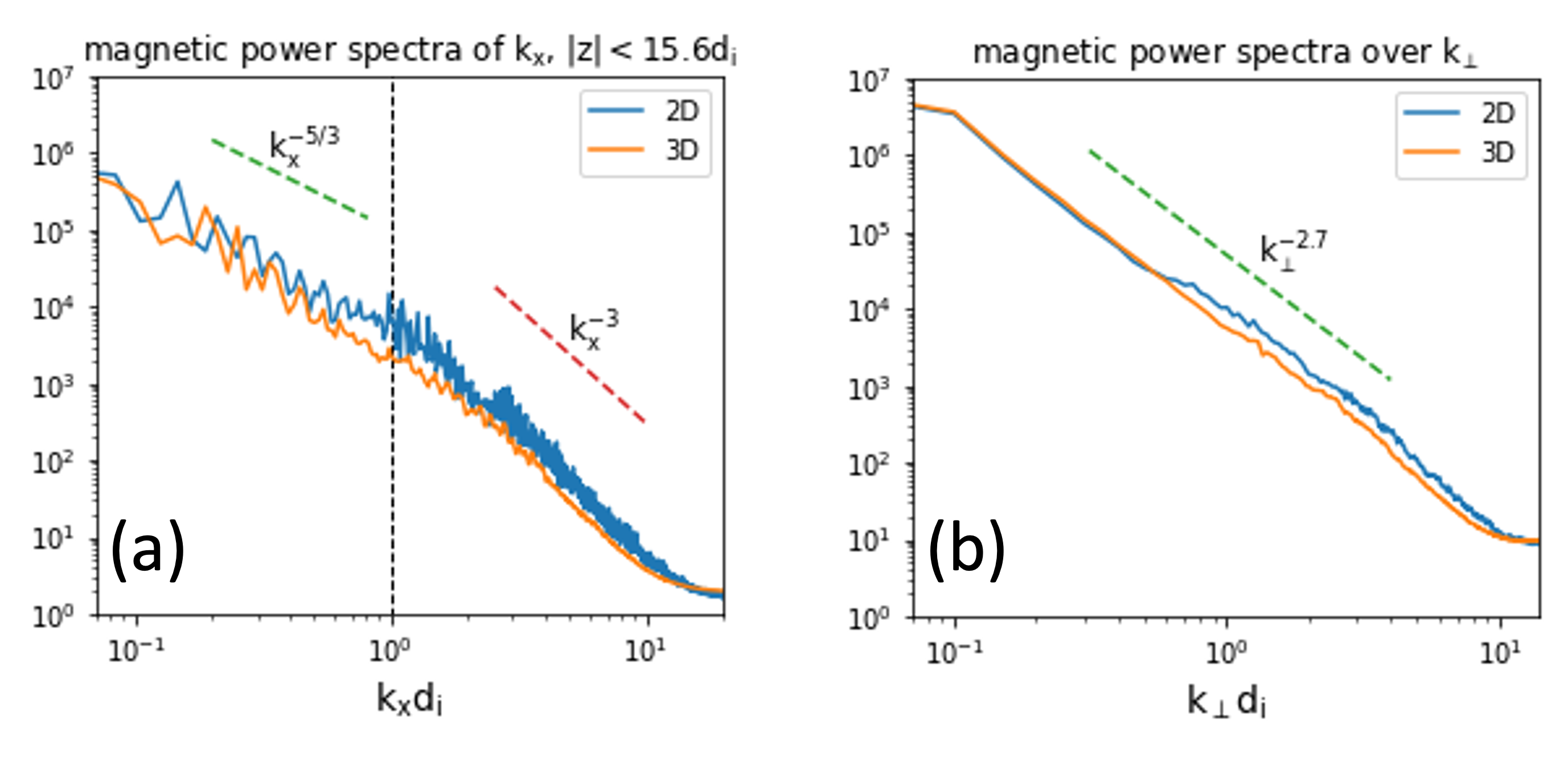} %figE.2.png}
\caption{(a) Magnetic power spectra over $k_x$ near the reconnection layer ($|z|<15.6d_i$) for Run 3 at $t\Omega_{ci}=200$ and for its 2D counterpart. Scalings $k_x^{-5/3}$ and $k_x^{-3}$ are plotted for reference at low and high wave numbers. (b) Magnetic power spectra over $k_\perp$ (perpendicular to the guide field direction) for Run 3 at $t\Omega_{ci}=200$ and for its 2D counterpart. A scaling $k_\perp^{-2.7}$ is plotted for reference. The 2D and 3D spectra are normalized to overlap and the vertical axes have arbitrary units. \label{fig6}}
\end{figure*}

%Note that emission intensity variations from solar flare remote-sensing observations can also indicate power-law power spectra with spectral indices between 1 and 2 \citep{Cheng2018}, but the uncertainty is too large that we will not discuss them in details here.
% Beyond ion scales, the difference between simulations and observations (2.7 and 5/3) is almost exactly one. This discrepency challenges our understanding of reconnection-driven turbulence and the particle acceleration process in the magnetotail. 
To better understand the spectra, we note that the MMS observations at the magnetotail probe the reconnection layer along the reconnection outflow direction, equivalently measuring spectra of $k_x$ in the layer. To have a more reasonable comparison, we examine the region near the reconnection layer ($|z|<15.6d_i$) to make spectra over $k_x$ and average them over $y$ and $z$ from our simulation Run 3 in Figure \ref{fig6}(a). This spectrum demonstrates power-law indices to be $\alpha\sim5/3$ below the $k_xd_i\sim1$ and $\alpha\sim3$ above it, in good agreement with the spectra measured by MMS \citep{Ergun2018,Ergun2020}. Therefore, magnetic power spectra at the magnetotail can be naturally driven by reconnection. For comparison, we also show the $k_\perp$ spectrum following \cite{Li2019b} in Figure \ref{fig6}(b) which again produces a different $\alpha\sim2.7$. 
%Previous studies of 3D reconnection also show that the $k_\perp$ spectral indices $p\sim2.7$ are similar for higher guide fields \citep{Daughton2014pop} and even for relativistic pair reconnection \citep{Guo2021apj}.

%The index around 5/3 does not necessarily indicate regular turbulence, but can suggest a reconnection-driven distribution of magnetic flux ropes.

We also show the results for corresponding 2D simulations in Figure \ref{fig6}(a-b). Surprisingly they are very similar to those for 3D simulations, even though the 2D simulations have no 3D turbulence with only laminar magnetic islands as the major magnetic structures. 
%In fact, even in the 3D simulations most magnetic energy is still stored in the $k_x-k_z$ plane ($k_y=0$) of the wave-vector space. 
Therefore, the index $\alpha\sim5/3$ in the $k_x$ spectra does not necessarily indicate Goldreich–Sridhar
 turbulence vortex cascade \citep{Ergun2018,Ergun2020}, but likely corresponds to a reconnection-driven size distribution of magnetic flux ropes or islands. 
 The $\alpha=5/3$ spectra steepening at scales smaller than $d_i$ is also consistent with the smallest flux ropes born in the kinetic reconnection current sheet. The largest scale in the $\alpha=5/3$ spectra around $k_xd_i\sim0.1$ also roughly corresponds to the largest flux-rope width $\sim60d_i$ in $x$ (see Figure \ref{fig5}). We also verify that in a smaller simulation Run 2 the $\alpha=5/3$ spectra have a shorter extension towards large scales (not shown) because of the smaller flux-rope size (due to the shorter reconnection time). Interestingly the higher guide field cases (those discussed in Section \ref{sec:higherBg} below) have very similar $k_x$  and $k_\perp$ spectral indices. We also verified that a longer $L_y$ up to $0.5L_x$ ($L_x=150d_i$ as in \cite{Li2019b}) does not significantly change the spectral indices, regardless of guide fields. We are preparing another upcoming paper that will quantitatively derive the island-size distribution and the associated power spectra consistent with the observed spectra. 
%we have also verified that the higher guide field cases (those discussed in Section \ref{sec:higherBg} below) have very similar $k_x$ (at mid-plane) and $k_\perp$ spectral indices. 
% \begin{figure}
%     \centering
%     \includegraphics[width=.6\textwidth]{fig6.5.png}
%     \caption{A schematic picture of propagating and growing islands in a reconnection layer.}
%     \label{fig6.5}
% \end{figure}
\section{reconnection with a higher guide field} \label{sec:higherBg}
\subsection{3D Dynamics}
\begin{figure*}
    \centering
\includegraphics[width=0.9\textwidth]{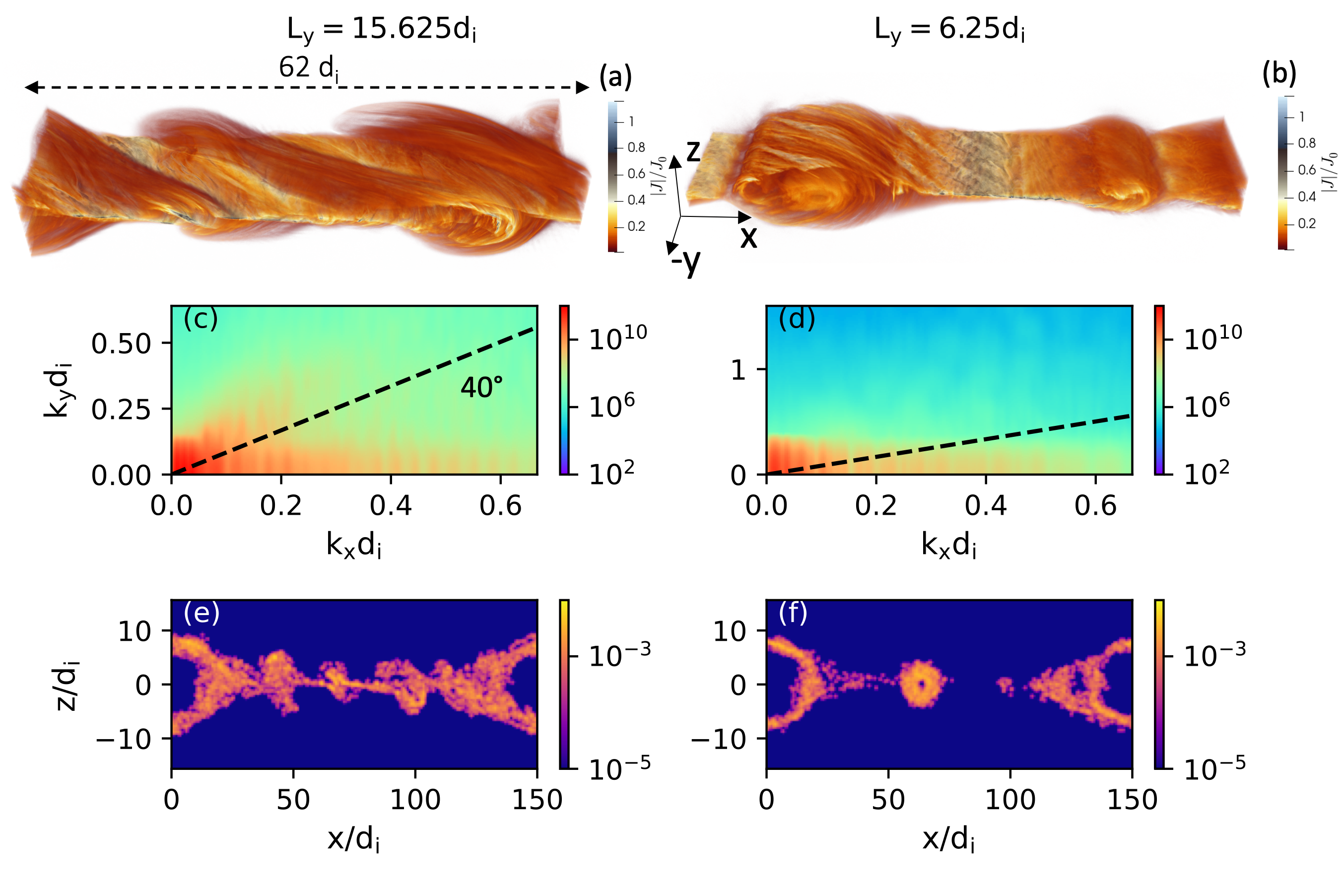} %figE.2.png}
\caption{Current density $|J|/en_0c$ for simulations Run 7 and 8 with different $y$ dimensions (a) $L_y=15.625d_i$ and (b) $L_y=6.25d_i$, $t\Omega_{ci}=125$, respectively. Panels (c) (d) show the corresponding Fourier analysis of $B_z$ over $k_x$ and $k_y$. An oblique angle of $40^\circ$ from $k_x$ is drawn in dash lines in both panels. Panels (e) (f) show the corresponding energetic electron density ($1.2 < \varepsilon/m_iV_A^2 < 2.4$) at $y=0$. \label{fig7}}
\end{figure*}

In the higher-guide-field regime, we find in our simulations that the flux ropes from oblique tearing modes can keep growing over time as they are advected with the reconnection outflows. Meanwhile, the flux ropes still maintain their oblique angles similar to the fastest growing linear tearing mode, even during the interaction between two resonant layers. This is possibly because the oblique orientation of reconnection x-lines at late stage is still controlled by a similar current filamentation tendency \citep{Liu2015} and the x-line orientation guides the orientation of flux ropes.
%We do not observe electron layer tearing mode as indicated by \cite{Daughton2011}, possibly because the flux ropes are oblique enough to avoid the strong current sheet between upstream fields and the flux ropes.  
Eventually the flux ropes become large and proportional to the system size. These overlapping oblique flux ropes of large sizes control the full 3D field-line chaos in reconnection of higher guide fields. We demonstrate this in Figure \ref{fig7} using simulations with $b_g=0.6$, which can trigger oblique tearing modes and still have relatively strong Fermi acceleration compared to $b_g>1$. To capture large oblique flux ropes in periodic domains of our simulations, the domain size in the guide-field direction $L_y$ needs to be large enough. Below we calculate the $L_y$ threshold. \cite{Liu2013prl} predict the maximum oblique angle of tearing modes $\theta_c=\arctan(1/b_g)$, equal to the oblique angle of the upstream magnetic fields. Numerically the fastest growing mode occurs roughly around $\theta_m\sim \theta_c/1.5$. The growth rates for $\theta>\theta_m$ drops quickly, so $\theta_m$ is also approximately the maximum angle with significant growth rates to provide the 3D effects. Thus, we focus on capturing large oblique flux ropes with this angle $\theta=\theta_m$ in our 3D domains. Imaging such a flux rope with size $D$ in the $x$ direction, from $y=0$ to $y=L_y$ the cross section of this flux rope needs to displace in $x$ for more than $D$ so that the flux rope does not ``bite" its tail to violate the periodic boundary condition. So, the $L_y$ threshold is 
%$L_{oblq}\sim1.5b_gD\propto b_gL_x$
\begin{equation}
    L_{oblq}=D/\tan(\theta_m)\sim
    D/\tan(\theta_c/1.5)\sim1.5D/\tan(\theta_c)=1.5b_gD.
\end{equation}
 In the above simulation $D\sim10d_i,b_g=0.6$, so $L_{oblq}\sim 9d_i$ with $\theta_m\sim40^{\circ}$. We show a simulation above this threshold in Figure \ref{fig7}(a) and another one below in Figure \ref{fig7}(b). The above-threshold case presents clear bifurcating oblique flux ropes above and below the reconnection layer. The below-threshold case contains mostly only quasi-2D straight flux ropes. We show the corresponding Fourier analysis of $B_z$ in Figure \ref{fig7}(c-d) in the space of $k_x$ and $k_y$. A Blackman window is applied along $z$ before the Fourier analysis to enforce a periodic boundary condition. The above-threshold case clearly has 3D oblique components with finite $k_y$, orienting at about $\theta_m=40^{\circ}$, which is consistent with the oblique angle of large oblique flux ropes estimated above. In contrast, the below-threshold case has only $k_x$, similar to a 2D simulation. We show the corresponding energetic electron density at $y=0$ in panel (e-f). The above-threshold case has energetic electrons spread throughout the reconnection layer, while the below-threshold case have them confined within the straight flux ropes. 

\subsection{Nonthermal Particle Acceleration}
\begin{figure*}
    \centering
\includegraphics[width=0.9\textwidth]{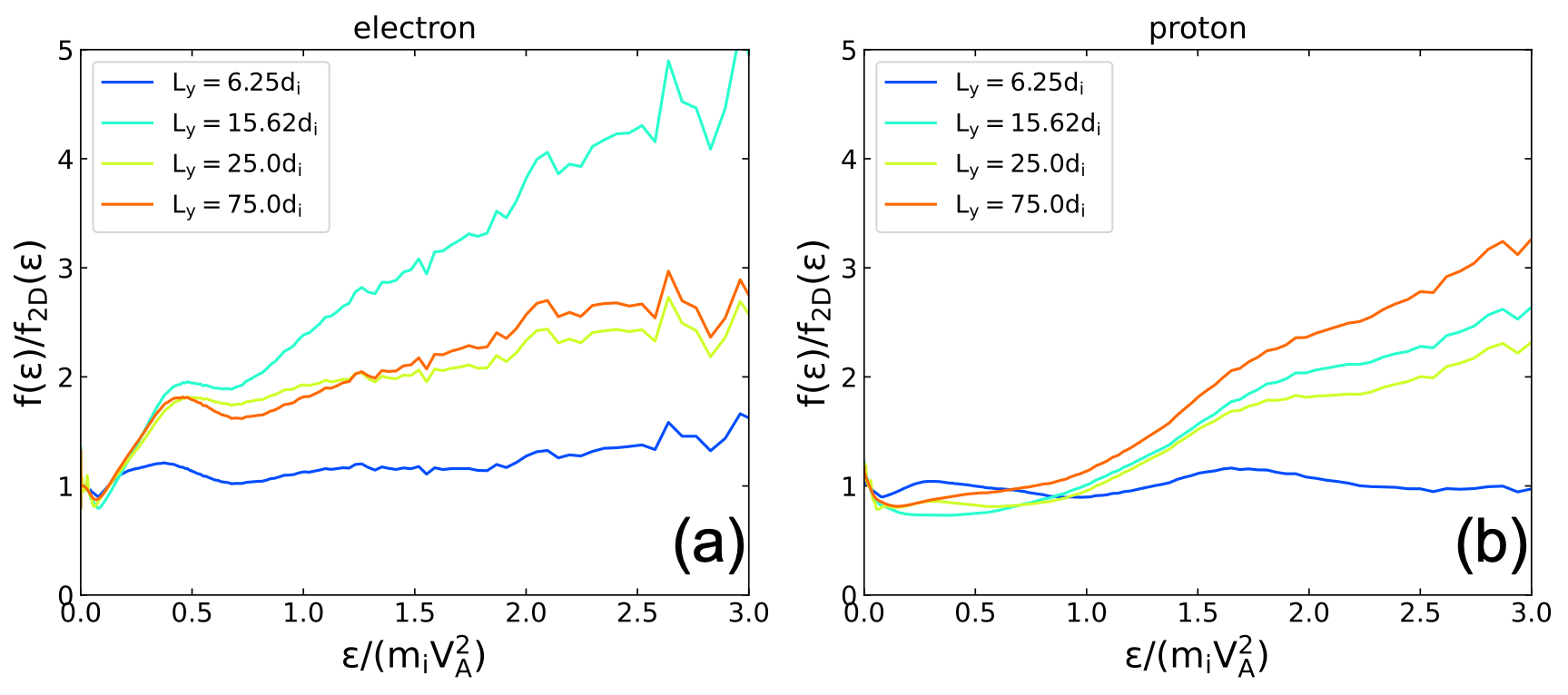} %figE.2.png}
\caption{Energy spectra of both species relative to 2D for simulations with different $L_y$ including Run 7-10. \label{fig8}}
\end{figure*}

These 3D effects with field-line chaos (and particle spreading) for above-threshold cases can lead to more efficient acceleration. Figure \ref{fig8} shows the increase in energy spectrum flux ($f_{3D}/f_{2D}$) for electrons and protons in simulations with different $L_y$ relative to the 2D counterpart, from below the threshold ($L_y=6.25$) to highly above the threshold ($L_y=75$). The below-threshold spectra are close to 2D ($f_{3D}/f_{2D}\sim1$) while the above-threshold cases achieve more energetic particles at high energies. Due to the higher guide fields with weaker Fermi acceleration, the acceleration enhancement in 3D is weaker than the low-guide-field regime \citep{Zhang2021}. Interestingly, the $L_y=15.62d_i$ case for electrons in cyan (but not obvious for ions) happens to have more acceleration than other 3D cases. We find in this run that there happens to be some more flux rope merging events locally in the middle of the reconnection layer, leading to more Fermi acceleration \citep{Drake2006} at the mergers and $E_\parallel$ acceleration at the merging x-lines. Electrons likely have much higher speed than ions and more 3D transport to make full use of the additional acceleration. Since these island mergers occur by chances, we would not consider this $L_y$ special. 

\begin{figure*}
    \centering
\includegraphics[width=0.9\textwidth]{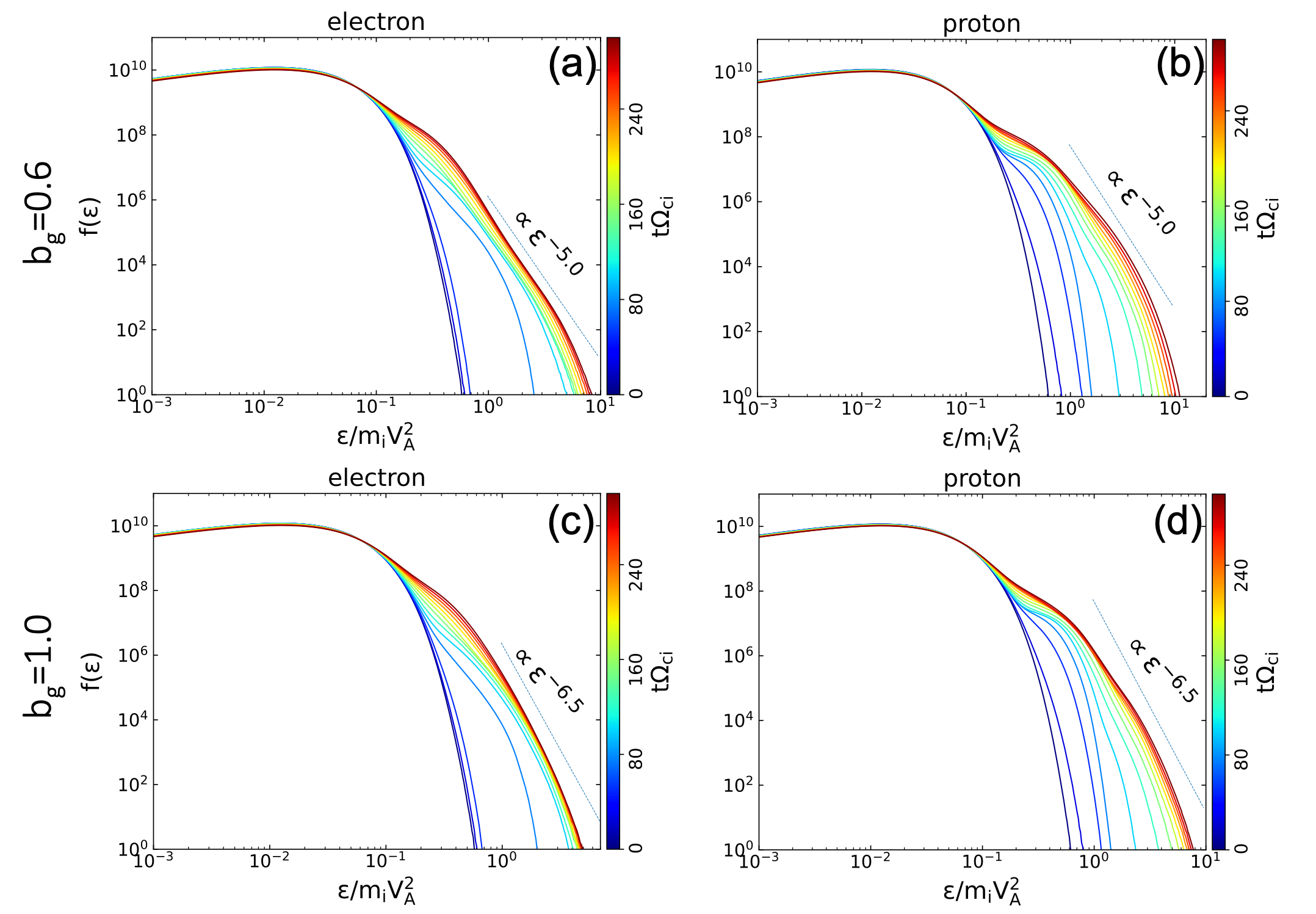} %figE.2.png}
\caption{Evolution of energy spectra for electrons and protons in large simulations Run 11 and 12, with guide fields 0.6 and 1 respectively.\label{fig9}}
\end{figure*}

We take advantage of this threshold in our domain-size design to perform large 3D simulations with unprecedented size with $L_x=300d_i$ (Run 11 and 12 in Table \ref{table1}). The upstream temperature is slightly higher than other simulations so that the grid size can be slightly larger (maintaining the ratio to the Debye length) to make the computational costs feasible. With more efficient and continuous acceleration in 3D, our 3D simulations for the first time accelerate both species into nonthermal power law spectra in the higher-guide-field regime. Figure \ref{fig9} shows the spectra for guide field $b_g = 0.6$ in panel (a-b) and guide field $b_g = 1.0$ in panel (c-d). Here the simulations run no more than 1 Alfv\'en crossing time (in contrast to 1.3 for low guide fields) because we find that higher guide fields hinder compression of the two large islands at the two ends of the reconnection layer, which enlarge their sizes to press back onto the reconnection layer at earlier times than low guide fields -- as an artifact of the periodic boundary in $x$. We apply Equation (\ref{p_formula}) from Fermi acceleration to the higher guide fields here by only changing $B_g/B_x\sim b_g$. While it predicts $p\sim4$ for low guide fields, it predicts $p\sim5$ for $b_g=0.6$ and $p\sim7$ for $b_g=1$. Figure \ref{fig9} shows electron and proton spectra reaching $p\sim5$ for $b_g=0.6$ and $p\sim6.5$ for $b_g=1$, roughly consistent with the prediction above from Fermi acceleration. The steeper spectra for higher guide fields result from the weaker Fermi acceleration. 
Note that near the end of simulations the high energy portions of the ion power-law spectra continue to extend to higher energy but the shoulders from injection shift to somewhat lower energies due to the reduction of upstream magnetic flux and thus $m_iV_A^2$ (getting 1.3 times lower). This leads to a slight distortion of the ion power-law spectra around the shoulders near the end of simulations. But this is just a limitation of the periodic boundaries over $z$ and it would not occur in reality with open incoming upstream flux.

\subsection{Oblique Flux-rope Kink Instability}
\begin{figure*}
    \centering
\includegraphics[width=0.7\textwidth]{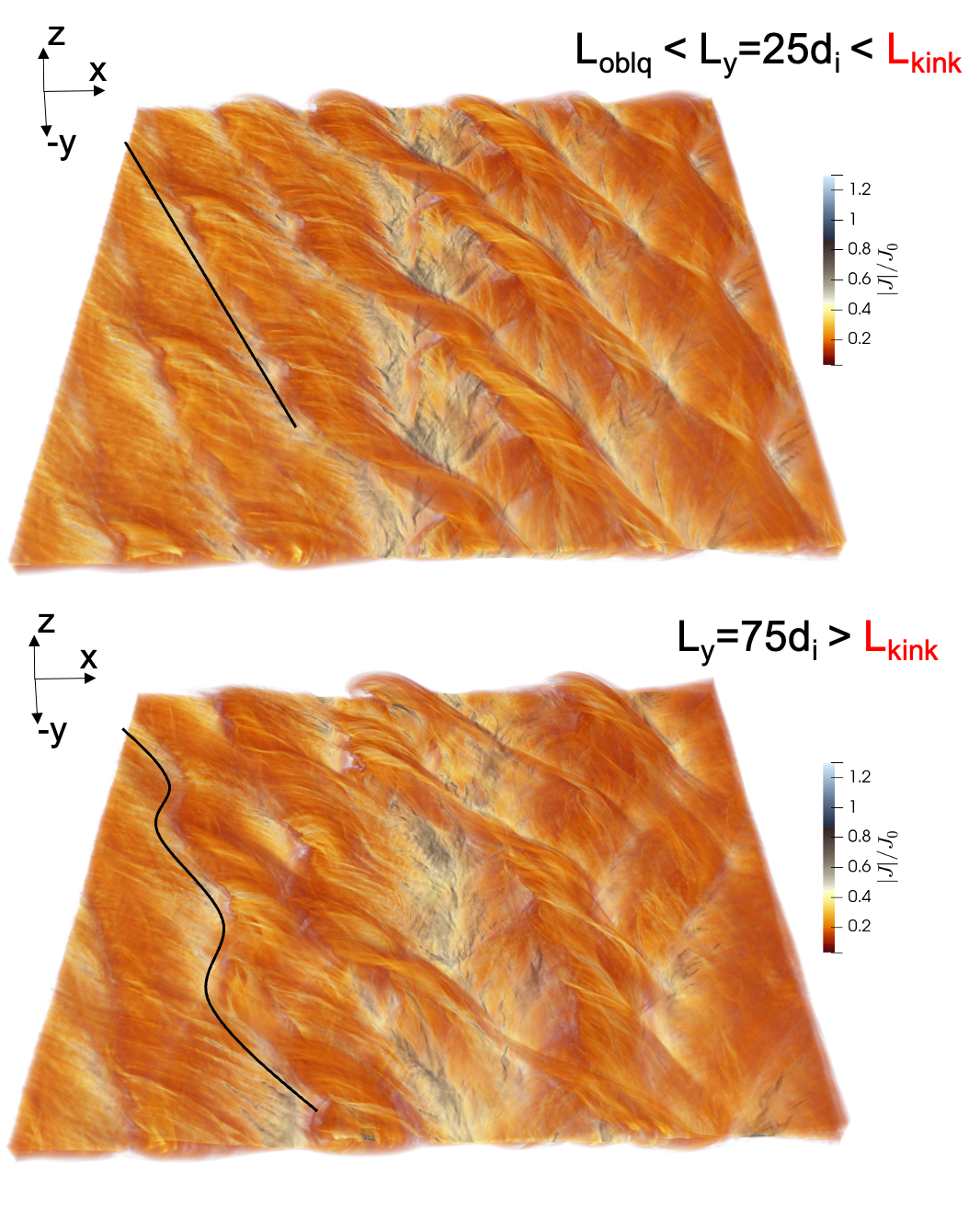} %figE.2.png}
\caption{Current density $|J|/en_0c$ for simulations Run 9 and 10 with different $L_y$, around $t\Omega_{ci}=80$. The relation to the $L_y$ thresholds $L_{oblq}$ (for large oblique flux ropes) and $L_{kink}$ (for oblique kink instability) is indicated. Two black lines near the flux rope edges emphasize their straight and kinking shapes.  \label{fig10}}
\end{figure*}

\begin{figure*}
    \centering
\includegraphics[width=0.5\textwidth]{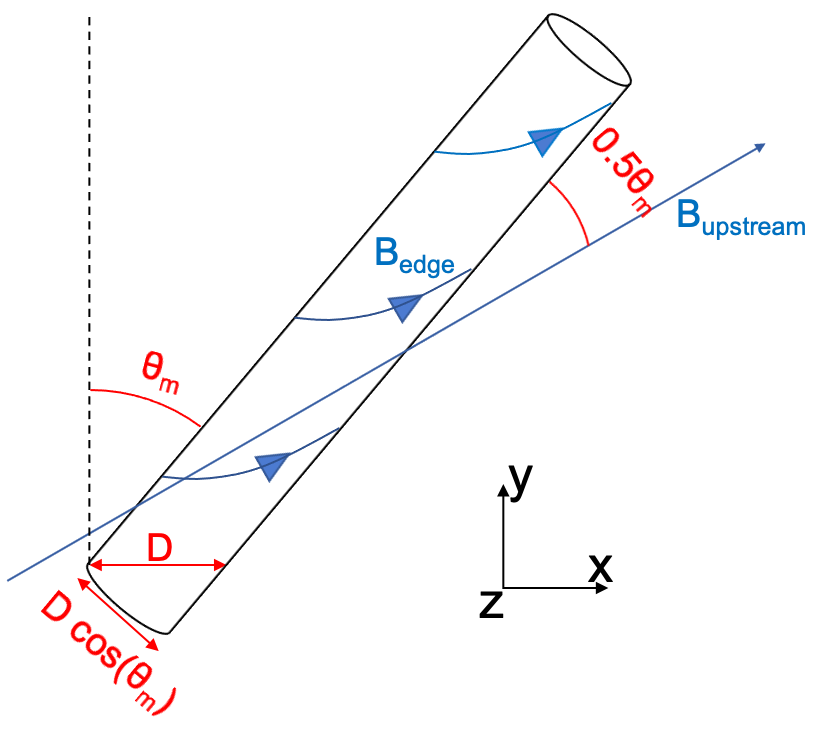} %figE.2.png}
\caption{A schematic picture that demonstrates the upstream magnetic fields wrapping around the edge of the oblique flux rope, in order to estimate the $L_y$ threshold $L_{kink}$ for the oblique kink instability. \label{fig11}}
\end{figure*}

We discover that the overlapping oblique flux ropes are not the only important process of the 3D dynamics with higher guide fields. The oblique flux ropes will run into the $m=1$ kink instability when $L_y$ exceeds a new threshold $L_{kink}$. For example, Figure \ref{fig10}(a-b) shows two above-$L_{oblq}$ simulations of $b_g=0.6$, below and above $L_{kink}\sim50d_i$ (to be calculated below) with $L_y=25$ and $75d_i$ respectively. The oblique flux ropes are formed and the reconnection bidirectional outflows in $x$ pull each flux ropes to both sides, resulting in some thinner elongated structures around the middle of the $x$ direction. With $L_y=25d_i$, 3D oblique flux ropes can exist but they are straight (indicated by the black line) and do not show extra $m=1$ kink dynamics. With $L_y=75d_i$, the oblique flux ropes are long enough to be $m=1$ kink unstable, driving extra exotic 3D kinking dynamics. We calculate $L_{kink}$ in the following. As demonstrated in Figure \ref{fig11}, consider an oblique flux rope with the angle $\theta_m$ and width $D$ in $x$. The cross section perpendicular to this oblique flux rope have a diameter $D\cos(\theta_m)$. The angle between the upstream magnetic field {$\mathbf{B_{upstream}}$} and the flux rope is $\theta_c-\theta_m\sim0.5\theta_m$. Since the upstream fields eventually wrap around the edge of the flux rope to become $\mathbf{B_{edge}}$, we can use the direction of the upstream field to calculate the minimum length of the flux rope to be $m=1$ kink unstable as $L_{FR}=\pi D \cos(\theta_m)/\tan(0.5\theta_m)$ (for the field to wrap around one circle). To contain this flux rope with periodic boundaries, a simulation needs to have a minimum $L_y$: 
\begin{equation}
    L_{kink}=L_{FR}*\cos(\theta_m)=\pi D \cos^2(\theta_m)/\tan(0.5\theta_m). 
\end{equation}
For $D=10d_i,b_g=0.6$, we get $L_{kink}\sim 50d_i$. This puts $L_y=75d_i$ above the threshold and $L_y=25d_i$ below the threshold, consistent with the simulations. Such oblique flux-rope kink instability drives extra 3D turbulent dynamics into the reconnection layer, which could potentially fragmentize the oblique flux ropes in a sufficiently large domain to turn the layer into a full turbulent state -- filled with fragmented kinking oblique flux ropes growing over time. An important question is whether this extra 3D turbulent dynamics can lead to extra acceleration. In Figure \ref{fig8}, the $L_y=75d_i$ case shows a little enhanced acceleration from the $L_y=25d_i$ case but it is insignificant, possibly because the overlapping oblique flux ropes with $L_y=25d_i$ already have field-line chaos sufficient for efficient acceleration. The effects of this extra instability will be further explored.

\section{Observation Implications}
%The magnetic power spectra, 3D flux-rope dynamics and spectral properties in our simulations should be compared to remote sensing and in-situ satellite observation of 3D reconnection.
The magnetic power spectra, 3D flux-rope dynamics and spectral properties in our simulations should have implications to remote sensing and in-situ satellite observation of 3D reconnection. The magnetic power spectra insensitive to guide field and 3D effects suggest a common power-spectral index $\alpha\sim 5/3$ (reflecting a common island-size distribution) measured along outflow directions for different reconnection-driven phenomena such as at the magnetotail, the magnetopause, heliospheric current sheets and solar flares. The 3D kink dynamics of flux ropes suggests the current density vectors in various flux ropes would swing in in-situ measurements. Specially with a guide field, the current density vectors in flux ropes at two sides of the reconnection midplane bifurcate to two distinct directions, with additional kinking swing around each direction. The downstream particle energy spectra should consist of a shoulder from injection and a power law from acceleration. The energy of the shoulder for ions $\sim m_iV_A^2$ is somewhat higher than electrons. The considerable guide-field dependence of power-law spectral indices (Equation \ref{p_formula}) can partly explain the wide range of observed indices \citep{Oka2018,Omodei2018,Bale2023nature}. The hardest spectral indices $p\sim4$ should be produced by low-guide-field low-beta reconnection such as in the magnetotail \citep{Ergun2020}, heliospheric current sheets \citep{Desai2022} and the impulsive phase of solar flares \citep{Chen2020}; reconnection with higher guide fields is expected to have softer indices.
\section{conclusion}
It has been a long-standing effort to understand the physics for nonthermal particle-acceleration in 3D magnetic reconnection, which shows contrasting dynamics for different guide fields. In this paper, we explored this by performing fully kinetic simulations with various parameters and domain sizes in both the weak and stronger guide field regimes. In both regime, we have uncovered the distinct 3D dynamics that leads to field-line chaos and efficient acceleration in reconnection and we have achieved nonthermal acceleration into power-law spectra. In the low-guide-field regime, the flux-rope kink instability enables the 3D dynamics for efficient acceleration. We took advantage of the flux-rope kink instability threshold to optimize our simulation domains to achieve the power-law spectra. The weak dependence of the spectra on the ion-to-electron mass ratio and $\beta$ ($\ll1$) implies that the plasma are magnetized enough by the magnetic flux ropes and adjacent exhausts for Fermi acceleration in our PIC simulations. While both electrons and protons are injected at reconnection exhausts, protons are majorly injected by perpendicular electric fields through Fermi reflections and electrons are injected by a combination of perpendicular and parallel electric fields. The magnetic power spectra suggest that reconnection can naturally drive the magnetic power spectra measured in the in-situ magnetotail observations. The power spectra may not indicate Goldreich–Sridhar turbulence vortex cascade but may instead reflect the size distribution of magnetic islands/flux ropes. In the higher-guide-field regime, the oblique flux ropes of large sizes control the domain-size threshold to capture the major 3D dynamics for efficient acceleration. We also made use of it to achieve nonthermal acceleration into power laws in energy spectra, with indices consistent with the prediction of  Fermi acceleration dependent on guide fields. Intriguingly, the oblique flux ropes can also run into flux-rope kink instability, driving extra exotic 3D dynamics for the reconnection layer that could potentially fragmentize the oblique flux ropes -- although it has not significantly further enhanced the acceleration. The effect of this instability on the 3D dynamics and particle acceleration will be further explored. 

This study also has some limitations. While our large-scale fully kinetic simulations have  domain sizes ($L_x\sim300d_i$) comparable to magnetospheric reconnection layers, they are still much smaller than macroscopic astrophysical systems such as solar flares. But the continuous Fermi acceleration and power law extension (shown in this paper and \cite{Zhang2021}) suggests that the power laws can continue to extend to much higher energy in macroscopic systems. The mass ratios in 3D simulations explored in this paper, though with a weak dependence, are still far from the realistic ratio due to the computational-cost constrains, which may be further investigated in the future. 

This work has broad implications for particle acceleration by 3D magnetic reconnection with a variety of guide fields, in not only heliosphysics (such as reconnection in the magnetosphere, solar winds and the solar corona), but also astrophysics (such as stellar flares and accretion disk flares) \citep{Ripperda2020,Nathanail2020}. 
\bigskip

% \begin{acknowledgements}
We gratefully acknowledge the helpful discussions in the SolFER DRIVE Science Center collaboration. Q.Z, F.G., W.D. and H.L. acknowledge the support from Los Alamos National Laboratory through the LDRD program and its Center
for Space and Earth Science (CSES), DOE OFES, and NASA programs through grant NNH17AE68I, 80HQTR20T0073, 80NSSC20K0627, 80HQTR21T0103 and 80HQTR21T0005, and through Astrophysical Theory Program. 
The simulations used resources provided by the Los Alamos National Laboratory Institutional Computing Program, the National Energy Research Scientific Computing Center (NERSC) and the Texas Advanced Computing Center (TACC). 

\bibliographystyle{aasjournal}

\begin{thebibliography}{}
\expandafter\ifx\csname natexlab\endcsname\relax\def\natexlab#1{#1}\fi
\providecommand{\url}[1]{\href{#1}{#1}}
\providecommand{\dodoi}[1]{doi:~\href{http://doi.org/#1}{\nolinkurl{#1}}}
\providecommand{\doeprint}[1]{\href{http://ascl.net/#1}{\nolinkurl{http://ascl.net/#1}}}
\providecommand{\doarXiv}[1]{\href{https://arxiv.org/abs/#1}{\nolinkurl{https://arxiv.org/abs/#1}}}

\bibitem[{{Arnold} {et~al.}(2021){Arnold}, {Drake}, {Swisdak}, {Guo}, {Dahlin},
  {Chen}, {Fleishman}, {Glesener}, {Kontar}, {Phan}, \& {Shen}}]{Arnold2021}
{Arnold}, H., {Drake}, J.~F., {Swisdak}, M., {et~al.} 2021, \prl, 126, 135101,
  \dodoi{10.1103/PhysRevLett.126.135101}

\bibitem[{{Bale} {et~al.}(2023){Bale}, {Drake}, {McManus}, {Desai}, {Badman},
  {Larson}, {Swisdak}, {Horbury}, {Raouafi}, {Phan}, {Velli}, {McComas},
  {Cohen}, {Mitchell}, {Panasenco}, \& {Kasper}}]{Bale2023nature}
{Bale}, S.~D., {Drake}, J.~F., {McManus}, M.~D., {et~al.} 2023, \nat, 618, 252,
  \dodoi{10.1038/s41586-023-05955-3}

\bibitem[{{Bowers} \& {Li}(2007)}]{Bowers2007}
{Bowers}, K., \& {Li}, H. 2007, \prl, 98, 035002,
  \dodoi{10.1103/PhysRevLett.98.035002}

\bibitem[{{Bowers} {et~al.}(2008){Bowers}, {Albright}, {Yin}, {Bergen}, \&
  {Kwan}}]{Bowers2008}
{Bowers}, K.~J., {Albright}, B.~J., {Yin}, L., {Bergen}, B., \& {Kwan},
  T.~J.~T. 2008, \pop, 15, 055703, \dodoi{10.1063/1.2840133}

\bibitem[{{Chen} {et~al.}(2020){Chen}, {Shen}, {Gary}, {Reeves}, {Fleishman},
  {Yu}, {Guo}, {Krucker}, {Lin}, {Nita}, \& {Kong}}]{Chen2020}
{Chen}, B., {Shen}, C., {Gary}, D.~E., {et~al.} 2020, Nature Astronomy, 4,
  1140, \dodoi{10.1038/s41550-020-1147-7}

\bibitem[{{Cheng} {et~al.}(2018){Cheng}, {Li}, {Wan}, {Ding}, {Chen}, {Zhang},
  \& {Liu}}]{Cheng2018}
{Cheng}, X., {Li}, Y., {Wan}, L.~F., {et~al.} 2018, \apj, 866, 64,
  \dodoi{10.3847/1538-4357/aadd16}

\bibitem[{{Cohen} {et~al.}(2020)}]{Cohen2020}
{Cohen}, C.~M.~S., {et~al.} 2020, \aap, \dodoi{10.1051/0004-6361/202039299}

\bibitem[{{Dahlburg} {et~al.}(1992){Dahlburg}, {Antiochos}, \&
  {Zang}}]{Dahlburg1992}
{Dahlburg}, R.~B., {Antiochos}, S.~K., \& {Zang}, T.~A. 1992, Physics of Fluids
  B, 4, 3902, \dodoi{10.1063/1.860347}

\bibitem[{Dahlin {et~al.}(2014)Dahlin, Drake, \& Swisdak}]{Dahlin2014}
Dahlin, J.~T., Drake, J.~F., \& Swisdak, M. 2014, \pop, 21,
  \dodoi{http://dx.doi.org/10.1063/1.4894484}

\bibitem[{{Dahlin} {et~al.}(2016){Dahlin}, {Drake}, \&
  {Swisdak}}]{Dahlin2016pop}
{Dahlin}, J.~T., {Drake}, J.~F., \& {Swisdak}, M. 2016, Physics of Plasmas, 23,
  120704, \dodoi{10.1063/1.4972082}

\bibitem[{Dahlin {et~al.}(2017)Dahlin, Drake, \& Swisdak}]{Dahlin2017}
Dahlin, J.~T., Drake, J.~F., \& Swisdak, M. 2017, \pop, 24, 92110,
  \dodoi{10.1063/1.4986211}

\bibitem[{{Daughton}(1998)}]{Daughton1998}
{Daughton}, W. 1998, \jgr, 103, 29429, \dodoi{10.1029/1998JA900028}

\bibitem[{{Daughton} {et~al.}(2014){Daughton}, {Nakamura}, {Karimabadi},
  {Roytershteyn}, \& {Loring}}]{Daughton2014pop}
{Daughton}, W., {Nakamura}, T.~K.~M., {Karimabadi}, H., {Roytershteyn}, V., \&
  {Loring}, B. 2014, Physics of Plasmas, 21, 052307, \dodoi{10.1063/1.4875730}

\bibitem[{{Daughton} {et~al.}(2011){Daughton}, {Roytershteyn}, {Karimabadi},
  {Yin}, {Albright}, {Bergen}, \& {Bowers}}]{Daughton2011}
{Daughton}, W., {Roytershteyn}, V., {Karimabadi}, H., {et~al.} 2011, Nature
  Physics, 7, 539, \dodoi{10.1038/nphys1965}

\bibitem[{{Desai} {et~al.}(2022){Desai}, {Mitchell}, {McComas}, {Drake},
  {Phan}, {Szalay}, {Roelof}, {Giacalone}, {Hill}, {Christian}, {Schwadron},
  {McNutt}, {Wiedenbeck}, {Joyce}, {Cohen}, {Davis}, {Krimigis}, {Leske},
  {Matthaeus}, {Malandraki}, {Mewaldt}, {Labrador}, {Stone}, {Bale},
  {Verniero}, {Rahmati}, {Whittlesey}, {Livi}, {Larson}, {Pulupa}, {MacDowall},
  {Niehof}, {Kasper}, \& {Horbury}}]{Desai2022}
{Desai}, M.~I., {Mitchell}, D.~G., {McComas}, D.~J., {et~al.} 2022, \apj, 927,
  62, \dodoi{10.3847/1538-4357/ac4961}

\bibitem[{Drake {et~al.}(2006)Drake, Swisdak, Che, \& Shay}]{Drake2006}
Drake, J.~F., Swisdak, M., Che, H., \& Shay, M.~A. 2006, Nature, 443, 553,
  \dodoi{10.1038/nature05116}

\bibitem[{{Ergun} {et~al.}(2018){Ergun}, {Goodrich}, {Wilder}, {Ahmadi},
  {Holmes}, {Eriksson}, {Stawarz}, {Nakamura}, {Genestreti}, {Hesse}, {Burch},
  {Torbert}, {Phan}, {Schwartz}, {Eastwood}, {Strangeway}, {Le Contel},
  {Russell}, {Argall}, {Lindqvist}, {Chen}, {Cassak}, {Giles}, {Dorelli},
  {Gershman}, {Leonard}, {Lavraud}, {Retino}, {Matthaeus}, \&
  {Vaivads}}]{Ergun2018}
{Ergun}, R.~E., {Goodrich}, K.~A., {Wilder}, F.~D., {et~al.} 2018, \grl, 45,
  3338, \dodoi{10.1002/2018GL076993}

\bibitem[{{Ergun} {et~al.}(2020){Ergun}, {Ahmadi}, {Kromyda}, {Schwartz},
  {Chasapis}, {Hoilijoki}, {Wilder}, {Stawarz}, {Goodrich}, {Turner}, {Cohen},
  {Bingham}, {Holmes}, {Nakamura}, {Pucci}, {Torbert}, {Burch}, {Lindqvist},
  {Strangeway}, {Le Contel}, \& {Giles}}]{Ergun2020}
{Ergun}, R.~E., {Ahmadi}, N., {Kromyda}, L., {et~al.} 2020, \apj, 898, 154,
  \dodoi{10.3847/1538-4357/ab9ab6}

\bibitem[{{French} {et~al.}(2019){French}, {Judge}, {Matthews}, \& {van
  Driel-Gesztelyi}}]{French2019}
{French}, R.~J., {Judge}, P.~G., {Matthews}, S.~A., \& {van Driel-Gesztelyi},
  L. 2019, \apjl, 887, L34, \dodoi{10.3847/2041-8213/ab5d34}

\bibitem[{Gary {et~al.}(2018)Gary, Chen, Dennis, Fleishman, Hurford, Krucker,
  McTiernan, Nita, Shih, White, \& Yu}]{Gary2018}
Gary, D.~E., Chen, B., Dennis, B.~R., {et~al.} 2018, \apj, 863, 83,
  \dodoi{10.3847/1538-4357/aad0ef}

\bibitem[{Guo {et~al.}(2014)Guo, Li, Daughton, \& Liu}]{Guo_2014}
Guo, F., Li, H., Daughton, W., \& Liu, Y.-H. 2014, Phys. Rev. Lett., 113,
  155005, \dodoi{10.1103/PhysRevLett.113.155005}

\bibitem[{{Guo} {et~al.}(2021){Guo}, {Li}, {Daughton}, {Li}, {Kilian}, {Liu},
  {Zhang}, \& {Zhang}}]{Guo2021apj}
{Guo}, F., {Li}, X., {Daughton}, W., {et~al.} 2021, \apj, 919, 111,
  \dodoi{10.3847/1538-4357/ac0918}

\bibitem[{{Guo} {et~al.}(2020){Guo}, {Liu}, {Li}, {Li}, {Daughton}, \&
  {Kilian}}]{Guo2020review}
{Guo}, F., {Liu}, Y.-H., {Li}, X., {et~al.} 2020, Physics of Plasmas, 27,
  080501, \dodoi{10.1063/5.0012094}

\bibitem[{Haggerty {et~al.}(2015)Haggerty, Shay, Drake, Phan, \&
  McHugh}]{Haggerty2015}
Haggerty, C.~C., Shay, M.~A., Drake, J.~F., Phan, T.~D., \& McHugh, C.~T. 2015,
  \grl, 42, 9657, \dodoi{10.1002/2015GL065961}

\bibitem[{{Johnson} {et~al.}(2022){Johnson}, {Kilian}, {Guo}, \&
  {Li}}]{Johnson2022}
{Johnson}, G., {Kilian}, P., {Guo}, F., \& {Li}, X. 2022, \apj, 933, 73,
  \dodoi{10.3847/1538-4357/ac7143}

\bibitem[{{Krucker} {et~al.}(2010){Krucker}, {Hudson}, {Glesener}, {White},
  {Masuda}, {Wuelser}, \& {Lin}}]{Krucker2010}
{Krucker}, S., {Hudson}, H.~S., {Glesener}, L., {et~al.} 2010, \apj, 714, 1108,
  \dodoi{10.1088/0004-637X/714/2/1108}

\bibitem[{{Le} {et~al.}(2009){Le}, {Egedal}, {Daughton}, {Fox}, \&
  {Katz}}]{Le2009}
{Le}, A., {Egedal}, J., {Daughton}, W., {Fox}, W., \& {Katz}, N. 2009, \prl,
  102, 085001, \dodoi{10.1103/PhysRevLett.102.085001}

\bibitem[{{Li} {et~al.}(2019{\natexlab{a}}){Li}, {Guo}, \& {Li}}]{Li2019a}
{Li}, X., {Guo}, F., \& {Li}, H. 2019{\natexlab{a}}, \apj, 879, 5,
  \dodoi{10.3847/1538-4357/ab223b}

\bibitem[{{Li} {et~al.}(2018){Li}, {Guo}, {Li}, \& {Birn}}]{Li2018}
{Li}, X., {Guo}, F., {Li}, H., \& {Birn}, J. 2018, \apj, 855, 80,
  \dodoi{10.3847/1538-4357/aaacd5}

\bibitem[{{Li} {et~al.}(2017){Li}, {Guo}, {Li}, \& {Li}}]{Li2017}
{Li}, X., {Guo}, F., {Li}, H., \& {Li}, G. 2017, \apj, 843, 21,
  \dodoi{10.3847/1538-4357/aa745e}

\bibitem[{{Li} {et~al.}(2019{\natexlab{b}}){Li}, {Guo}, {Li}, {Stanier}, \&
  {Kilian}}]{Li2019b}
{Li}, X., {Guo}, F., {Li}, H., {Stanier}, A., \& {Kilian}, P.
  2019{\natexlab{b}}, \apj, 884, 118, \dodoi{10.3847/1538-4357/ab4268}

\bibitem[{{Li} {et~al.}(2021){Li}, {Guo}, \& {Liu}}]{Li2021review}
{Li}, X., {Guo}, F., \& {Liu}, Y.-H. 2021, Physics of Plasmas, 28, 052905,
  \dodoi{10.1063/5.0047644}

\bibitem[{Lin(2011)}]{Lin2011}
Lin, R.~P. 2011, \ssr, 159, 421, \dodoi{10.1007/s11214-011-9801-0}

\bibitem[{Liu {et~al.}(2011)Liu, Li, Yin, Albright, Bowers, \&
  Liang}]{WLiu2011}
Liu, W., Li, H., Yin, L., {et~al.} 2011, Physics of Plasmas, 18, 052105,
  \dodoi{10.1063/1.3589304}

\bibitem[{{Liu} {et~al.}(2013){Liu}, {Daughton}, {Karimabadi}, {Li}, \&
  {Roytershteyn}}]{Liu2013prl}
{Liu}, Y.-H., {Daughton}, W., {Karimabadi}, H., {Li}, H., \& {Roytershteyn}, V.
  2013, \prl, 110, 265004, \dodoi{10.1103/PhysRevLett.110.265004}

\bibitem[{Liu {et~al.}(2015)Liu, Hesse, \& Kuznetsova}]{Liu2015}
Liu, Y.-H., Hesse, M., \& Kuznetsova, M. 2015, Journal of Geophysical Research:
  Space Physics, 120, 7331, \dodoi{https://doi.org/10.1002/2015JA021324}

\bibitem[{{Nathanail} {et~al.}(2020){Nathanail}, {Fromm}, {Porth}, {Olivares},
  {Younsi}, {Mizuno}, \& {Rezzolla}}]{Nathanail2020}
{Nathanail}, A., {Fromm}, C.~M., {Porth}, O., {et~al.} 2020, \mnras, 495, 1549,
  \dodoi{10.1093/mnras/staa1165}

\bibitem[{Oka {et~al.}(2015)Oka, Krucker, Hudson, \& Saint-Hilaire}]{Oka2015}
Oka, M., Krucker, S., Hudson, H.~S., \& Saint-Hilaire, P. 2015, \apj, 799, 129.
\newblock \url{http://stacks.iop.org/0004-637X/799/i=2/a=129}

\bibitem[{{Oka} {et~al.}(2018){Oka}, {Birn}, {Battaglia}, {Chaston}, {Hatch},
  {Livadiotis}, {Imada}, {Miyoshi}, {Kuhar}, {Effenberger}, {Eriksson},
  {Khotyaintsev}, \& {Retin{\`o}}}]{Oka2018}
{Oka}, M., {Birn}, J., {Battaglia}, M., {et~al.} 2018, \ssr, 214, 82,
  \dodoi{10.1007/s11214-018-0515-4}

\bibitem[{{Oka} {et~al.}(2023){Oka}, {Birn}, {Egedal}, {Guo}, {Ergun},
  {Turner}, {Khotyaintsev}, {Hwang}, {Cohen}, \& {Drake}}]{Oka2023SSRv}
{Oka}, M., {Birn}, J., {Egedal}, J., {et~al.} 2023, \ssr, 219, 75,
  \dodoi{10.1007/s11214-023-01011-8}

\bibitem[{Omodei {et~al.}(2018)Omodei, Pesce-Rollins, Longo, Allafort, \&
  Krucker}]{Omodei2018}
Omodei, N., Pesce-Rollins, M., Longo, F., Allafort, A., \& Krucker, S. 2018,
  \apj, 865, L7, \dodoi{10.3847/2041-8213/aae077}

\bibitem[{{Onofri} {et~al.}(2006){Onofri}, {Isliker}, \& {Vlahos}}]{Onofri2006}
{Onofri}, M., {Isliker}, H., \& {Vlahos}, L. 2006, \prl, 96, 151102,
  \dodoi{10.1103/PhysRevLett.96.151102}

\bibitem[{{Oz} {et~al.}(2011){Oz}, {Myers}, {Yamada}, {Ji}, {Kulsrud}, \&
  {Xie}}]{Oz2011}
{Oz}, E., {Myers}, C.~E., {Yamada}, M., {et~al.} 2011, \pop, 18, 102107,
  \dodoi{10.1063/1.3647567}

\bibitem[{{Richard} {et~al.}(2024){Richard}, {Sorriso-Valvo}, {Yordanova},
  {Graham}, \& {Khotyaintsev}}]{Richard2024prl}
{Richard}, L., {Sorriso-Valvo}, L., {Yordanova}, E., {Graham}, D.~B., \&
  {Khotyaintsev}, Y.~V. 2024, \prl, 132, 105201,
  \dodoi{10.1103/PhysRevLett.132.105201}

\bibitem[{{Ripperda} {et~al.}(2020){Ripperda}, {Bacchini}, \&
  {Philippov}}]{Ripperda2020}
{Ripperda}, B., {Bacchini}, F., \& {Philippov}, A.~A. 2020, \apj, 900, 100,
  \dodoi{10.3847/1538-4357/ababab}

\bibitem[{{Shih} {et~al.}(2009){Shih}, {Lin}, \& {Smith}}]{Shih2009}
{Shih}, A.~Y., {Lin}, R.~P., \& {Smith}, D.~M. 2009, \apjl, 698, L152,
  \dodoi{10.1088/0004-637X/698/2/L152}

\bibitem[{Yamada {et~al.}(2010)Yamada, Kulsrud, \& Ji}]{Yamada2010}
Yamada, M., Kulsrud, R., \& Ji, H. 2010, Rev. Mod. Phys., 82, 603,
  \dodoi{10.1103/RevModPhys.82.603}

\bibitem[{{Zenitani} \& {Hoshino}(2005)}]{Zenitani2005}
{Zenitani}, S., \& {Hoshino}, M. 2005, \apjl, 618, L111, \dodoi{10.1086/427873}

\bibitem[{{Zhang} {et~al.}(2019{\natexlab{a}}){Zhang}, {Drake}, \&
  {Swisdak}}]{Zhang2019}
{Zhang}, Q., {Drake}, J.~F., \& {Swisdak}, M. 2019{\natexlab{a}}, \pop, 26,
  072115, \dodoi{10.1063/1.5104352}

\bibitem[{{Zhang} {et~al.}(2019{\natexlab{b}}){Zhang}, {Drake}, \&
  {Swisdak}}]{Zhang2019b}
{Zhang}, Q., {Drake}, J.~F., \& {Swisdak}, M. 2019{\natexlab{b}}, \pop, 26, 102115, \dodoi{10.1063/1.5121782}

\bibitem[{Zhang {et~al.}(2024)Zhang, Guo, Daughton, Li, Le, Phan, \&
  Desai}]{Zhang2024prl}
Zhang, Q., Guo, F., Daughton, W., {et~al.} 2024, Phys. Rev. Lett., 132, 115201,
  \dodoi{10.1103/PhysRevLett.132.115201}

\bibitem[{Zhang {et~al.}(2021)Zhang, Guo, Daughton, Li, \& Li}]{Zhang2021}
Zhang, Q., Guo, F., Daughton, W., Li, H., \& Li, X. 2021, Phys. Rev. Lett.,
  127, 185101, \dodoi{10.1103/PhysRevLett.127.185101}

\end{thebibliography}



\end{document}